\documentclass[aps,preprint,superscriptaddress,amsmath]{revtex4}

\usepackage[dvips]{graphicx}
\usepackage{dcolumn}
\usepackage{amsmath}
\usepackage{threeparttable}
\usepackage{appendix}
\usepackage{bm}

\newcommand{\be}{\begin{equation}}
\newcommand{\ee}{\end{equation}}
\newcommand{\bea}{\begin{eqnarray}}
\newcommand{\eea}{\end{eqnarray}}
\newcommand{\bean}{\begin{eqnarray*}}
\newcommand{\eean}{\end{eqnarray*}}

\begin{document}



\title{Systematically convergent method for accurate total energy calculations with localized atomic orbitals}

\author{S. Azadi}
\email{azadi@sissa.it}
\affiliation{ SISSA, International School for Advanced Studies, 34151, Trieste, Italy}
\author{C. Cavazzoni}
\affiliation{CINECA, 40033, Bologna, Italy}
\author{S. Sorella}
\email{sorella@sissa.it}
\affiliation{ SISSA, International School for Advanced Studies, 34151, Trieste, Italy}
\affiliation{ DEMOCRITOS Simulation Center
CNR-IOM Istituto Officina dei Materiali, 34151, Trieste, Italy}


\date{\today}

\begin{abstract} 

We introduce a method for solving a self consistent electronic calculation within  localized atomic orbitals,  that  allows us to converge to the complete basis set (CBS) limit in a stable, controlled, and systematic way.
We compare our results with the ones obtained with a standard quantum chemistry package for the  simple benzene molecule. We find perfect agreement for small basis set and  show that, within our scheme, it is possible to work with a very large basis in an efficient and stable way. Therefore we can avoid to introduce any extrapolation to reach the CBS limit.

In our study we have also carried out variational Monte Carlo (VMC) and lattice regularized 
diffusion Monte Carlo (LRDMC) with a standard  many-body wave function (WF) defined by
the product of a Slater determinant and a Jastrow factor. 
Once  the Jastrow factor  is  optimized by keeping fixed the Slater determinant provided by our new scheme, we obtain  
a very good description of the atomization energy of the benzene molecule 
only when the  basis of atomic orbitals is large enough and 
close to the CBS limit, yielding the lowest variational energies. 
\end{abstract}

\maketitle

\section{Introduction}
\label{introduction}
 The use of localized basis sets is becoming more and more important in several 
electronic structure methods, because it allows a dramatic reduction of the 
dimension $N$ of the single particle basis, i.e.  much smaller than 
standard plane wave basis set dimensions.
For instance in quantum chemistry calculations it is very difficult, if not impossible, to use  post Hartree-Fock methods  with extremely large basis sets, 
because their computational cost scales with a large power of $N$ ($N^4 -N^7$).
On the other hand all linear scaling methods \cite{Rayson,Gulans,Cervera,Wang} are based on a suitably 
localized  basis and exploit the fact that the matrix elements of the 
Hamiltonian decay very rapidly with the distance among the 
orbital centers. Several other applications are known, and it is basically impossible to list all of them. 

Although our findings apply generally to all the above issues, 
in this paper we focus in particular  on the use of a localized basis set 
for Monte Carlo optimization of wave functions because the 
number of variational parameters -e.g. in a Slater determinant- 
is proportional to $N$, and only with a localized basis it remains in 
 a reasonable range even for large systems. Therefore, thanks 
to the recent advances in the optimization techniques in quantum Monte 
Carlo calculations \cite{Mari,Umrigar1,Umrigar2,Casula}, it is possible nowadays 
to optimize a full many-body wave function with  
 an affordable computational time (proportional to $N^3-N^4$).

Despite the use of localized basis sets is becoming more and more popular 
there is an important issue on how to converge 
to the so called ''complete basis set'' (CBS) limit. 
In plane wave calculations this is usually 
achieved by systematically increasing the kinetic energy 
cutoff so that  a reliable and very well controlled convergence is possible. 
Unfortunately  within a localized basis set framework,
until now it is not possible to reach  the same level of accuracy 
 of the plane wave approach, 
because a too large value of $N$ leads to redundancy of 
the basis and typically to numerical instabilities.
On the other hand, working with a small basis far from the CBS limit, 
leads to the well known basis superposition error, that deteriorates
the accuracy in the description of the chemical bond.

The difficulty to work with a localized basis of $N$ non-orthogonal orbitals 
$\phi_i(\vec r)$  ($i=1,\cdots, N$) can be quantified by considering the $N \times N$ overlap matrix:
\begin{equation} \label{sov}
S_{i,j} = \langle \phi_i| \phi_j \rangle=\int d^3\vec r  \phi_i(\vec r) \phi_j (\vec r ). 
\end{equation}
In the following we assume that the atomic orbitals are normalized so that the 
diagonal elements of the above overlap matrix are identically equal to one.
Since this is an overlap matrix it is also positive definite, 
namely all its eigenvalues $s_i$, henceforth assumed in ascending order,
are positive $s_i\ge0$, and  a vanishing eigenvalue occurs only if the 
orbitals are linearly dependent.

If the condition number of this positive definite matrix, namely   
the ratio $s_{cond}={s_N \over s_1}$ 
between the largest  $s_N$
($s_{N} <N$, as the trace of S is  $\sum_i S_{i,i}= \sum_i s_i =N$) 
 and the lowest  eigenvalue  $s_1$
is below 
a certain threshold, 
calculations are very difficult and the convergence to the 
CBS limit is impossible to reach with standard methods.
This problem has been often circumvented by extrapolation procedures and 
/or relying on cancellation errors \cite{Glendening,Maslen,Weck,Simmonett}.
However we believe that, at least for a simple self-consistent field (SCF) calculation
 based on the density functional theory (DFT) within the standard local density approximation (LDA), it is very important to 
converge to the CBS limit, essentially in the same way as in a plane wave calculation.

Here we show that this important task can be achieved by applying 
a simple strategy for the diagonalization of the Hamiltonian, strategy  that can be applied to any SCF calculation.
Moreover we find that, even for the atomization energy 
of the simple benzene molecule, it is necessary to work close  to the CBS limit 
in order to obtain a well converged  result because,  at least in this case,  
 energy differences approach the CBS limit  only slightly faster than the total energy.

In this work, as already mentioned, we are interested to combine  
the Slater determinant obtained by an SCF calculation 
with a so called Jastrow factor, also expanded in a localized 
basis $\phi_i(\vec r)$. In this way an  accurate many-body 
wave function is defined, that typically describes rather well the electron 
correlation, often much better than the original SCF calculation. 
Indeed in our  QMC calculations, 
 we have also found that it is very important to use a very large 
basis set for the DFT Kohn-Sham orbitals, 
because only in this case the corresponding Slater 
determinant is very close to the optimal one, namely the one that minimizes 
the energy in presence of the Jastrow factor. 
After that we obtain a very accurate atomization energy for this simple 
molecule, that is compatible with the experiments, despite our strong 
restriction of the variational ansatz.

The paper is organized as follows. In Sec.II we describe the basis of 
localized atomic orbitals and the variational wave function used in this work,  
we briefly review the 
SCF LDA method 
and we show how to  work with a large basis of localized atomic orbitals.  
In Sec.III we present our results for the     benzene molecule, and finally 
in Sec. IV we draw our conclusions.

\section{Computational details}
\subsection{Localized basis set}

In our DFT LDA-based calculations, we have used Slater (S) exchange and Perdew \cite{perdew81} (P) correlation functionals. 
The Kohn-Sham (KS) equations are solved by expanding the electronic orbitals in a Gaussian type orbital (GTO) basis set. 
Only four valence electrons are taken into account for the  carbon atom. The 1s core electrons are considered by atomic 
pseudo-potentials(PPs), that are also used for the hydrogen atom in order to eliminate the divergent electron-ion attraction at short distance \cite{filippipseudo}. 

In the test  calculation presented here, 
we consider the benzene molecule with  experimental carbon and hydrogen 
atomic positions.
We define a localized basis set centered on each atom by using simple
Gaussians $ exp ( -Z_{i} r^2 ) $  with given angular momentum: $s,p,d,f,g,...$.
In order to achieve convergence in a systematic way we define the  standard 
even-tempered sequence for GTO exponents  $Z_i$:
\begin{equation}
Z_i = \alpha \beta^{i}
\end{equation}
for $i=0,\cdots n_l-1$ with $\alpha=Z_{min}$ and  $\alpha \beta^{n_l} = Z_{max}$, where $n_l$ is the number of exponents used for the angular momentum $l$.
The maximum number $n$ of exponents is used only on the s-wave channel $n_0=n$,
whereas for the higher angular momenta $l=1,2,3,4,...$  the number of exponents $n_l$ is smaller and is given  by following choice:
\begin{equation}
n_l = n_0 -2l 
\end{equation}
Notice that the value of $\beta$ is implicitly defined 
by the choice of $Z_{max}$ that will be discussed in the following.

This basis is obviously complete as long as  
  $n$ and the maximum $l$ ($ l \le l_{max}$) are systematically 
increased until convergence is reached. 
The obvious advantage of our even-tempered GTO set is that the exponent 
sequence 
is determined by only two parameters, $Z_{max}$ and $Z_{min}$. 
Our purpose is indeed to show that a systematic convergence with $n$ and 
$l_{max} \le 4$ can be obtained by using a large but not prohibitive value
of $n$.  
In practice for large $n$, the SCF energy is almost independent on $Z_{max}$ and $Z_{min}$ and therefore the choice of these two parameters 
can be done  by optimizing the DFT energy for few test cases, 
and by checking the stability of these two parameter values for large $n$.
As a result we have verified that the simple choice  $Z_{min}=0.2$ and
$Z_{max}=10$ is nearly optimal for all $n$ in the benzene case, as well 
as for the carbon atom. 
We have therefore adopted these two parameter values in all the forthcoming 
calculations.
Probably by optimizing all exponents much faster convergence can be obtained 
but in this work we want to emphasize that it is possible to work with a large 
basis set, deliberately larger than necessary because not fully optimized,  
and obtain accurate and systematically converging  results without 
limitations or constraints, likewise a plane-wave based approach.
As a result we show that we can achieve this task by using $n$ as large as 
$\lesssim 30$, that represents a very important restriction of the dimension 
$N$ of the basis, roughly  two or three orders of magnitude smaller compared 
to a plane wave based approach with the same accuracy in the total energy, 
e.g.  an accuracy of $0.1 mH$ in the total energy of the benzene molecule 
can be achieved with $N\simeq1000$ localized orbitals, 
whereas  at least $200^3$ plane-waves are necessary for the same target.  

\subsection{ Description of the SCF method}
The DFT functional, within the LDA approximation,  can be evaluated on 
 a given set of 
atomic orbitals $\phi_i( \vec r)$, and is then defined by the 
overlap matrix $S$ in Eq.(\ref{sov}) 
and the one body Hamiltonian matrix elements: 
\begin{equation}
H_{i,j} =  \langle \phi_i | H^{1b} | \phi_j \rangle +
  \langle \phi_i | v_H + v_{xc}  | \phi_j \rangle
\end{equation} 
where $H^{1b}$ contains Kinetic energy, electron-ion interaction, and 
the pseudopotentials used, whereas $v_H$ and $v_{xc}$ 
 are the Hartree and the exchange and correlation potentials, respectively,
 both defined only by the total electronic density $n (\vec r)$. 
In this non-orthonormal basis
the Kohn-Sham orbitals are given by: 
\begin{equation} \label{defpsi}
\psi_i( \vec r ) = \sum_j \chi_{ij} \phi_j( \vec r) 
\end{equation}
where the coefficients $\chi_{i,j}$ can be obtained by solving the 
generalized eigenvalue problem:
\begin{equation} \label{scfed}
\sum\limits_{j} H_{i,j} \chi_{i,j} = E_i \sum\limits_j S_{i,j} \chi_{i,j} 
\end{equation} 
The density is  in turn defined by occupying the Kohn-Sham orbitals 
up to the Fermi energy  $E_F$ 
\begin{equation}
n(\vec r) = \sum_{E_i \le E_F} 2 \psi_i^*(\vec r)  \psi_i( \vec r)
\end{equation}
and self consistency is reached when the output 
density obtained after the diagonalization coincides within numerical 
accuracy with the input density used to evaluate $v_H$ and $v_{xc}$.

In recent years computer performances have substantially  increased and 
the calculation of overlap and Hamiltonian matrices elements of the 
above types can be done rather efficiently by straightforward  integration 
over a mesh, e.g.  by replacing the integrals with appropriate summations over a set of 
electronic positions $\vec r_I $ uniformly distributed in a finite volume 
 $V=L_x\times L_y \times L_z$, spanned by $ n_x \times n_y \times n_z $
 mesh points referred to the $x,y,z$ Cartesian axes respectively, namely:
\begin{eqnarray}
S_{i,j} &= &  v \sum\limits_I \phi_i(\vec r_I) \phi_j(\vec r_I) \nonumber \\ 
H_{i,j} &= & { 1 \over 2 } 
v \sum \limits_I \left[ \phi_i  ( \vec r_I )   (H^{KS}   \phi_j ( \vec r_I )) 
 + ( H^{KS} \phi_i (\vec r_I)) \phi_j ( \vec r_I )  \right] 
\end{eqnarray}
where  $H^{KS}= H^{1b}+ v_H + v_{xc} $ is the Kohn-Sham Hamiltonian and 
$v = { L_x L_y L_z \over n_x n_y n_z} $ is 
the elementary box volume of the mesh grid.
Notice that in the above discretization of the integrals we have symmetrized 
the Hamiltonian matrix elements, thus restoring the Hermitian property  of the Hamiltonian ($H_{i,j}=H_{j,i}$) even for a finite mesh.

Further  important improvements have been introduced   to allow a  
fast and efficient convergence of these matrix elements by 
increasing the box volume and the size of the mesh ($n_x, n_y, n_z \to \infty$).
For instance  the origin of the mesh grid was chosen  in a way to 
maximize the minimum distance between the mesh and the ion positions.
Moreover the Hartree potential in atomic units was calculated
 by the convolution: 
\begin{equation} \label{fftconv}
v_H(\vec r_J)  =v  \sum\limits_{ I \ne J} 
   1/|\vec r_I - \vec r_J|  n (\vec r_I)  + C n( \vec r_J) = 1/V^A \sum_q  v^c_q n_{-q} 
\end{equation}
where $n_q$ and $v^c_q$  are the density and the Coulomb potential 
Fourier transform, respectively, whereas the finite constant $C$ takes into
 account the infinite contribution of the Coulomb potential for $\vec r_I = \vec r_J$, in a way that will be discussed later on.
For open systems, the 
 convolution was computed on a box of twice linear dimension 
with volume $V^A= 8 V$ and with the origin $\vec r = (0,0,0)$ at the center of the 
box,  namely $n_x \to 2 n_x, n_y \to 2 n_y, n_z \to 2 n_z $ and 
$ |x|\le L_x, |y| \le L_y, |z| \le L_z$. 
The convoluted density $n(\vec r)$ is assumed to vanish in $V^A$ outside the physical volume $V$ ( $0 \le x \le L_x , 0 \le  y \le L_y , 0 \le z \le L_z$) because the charge density is decaying exponentially fast at large distance from the atoms. Then it is defined periodic with period $(2L_x,2L_y,2L_z)$, so that wavevectors $q$ are correspondingly quantized  $q= (  \pi/ L_x n, \pi/L_y m , \pi/L_z l )$ with integers $n,m,l$, and  the standard 
and extremely efficient convolution algorithm based on fast Fourier transform 
can be applied, with $v^c_q = 
 \sum\limits_{ r_J  \in V^A, r_J \ne 0} v/|\vec r_J | e^{ -i q r_J }+C$.
By using this enlarged box $V^A$,   
we can avoid to add fictitious contributions to the Hartree potential, 
that would have  raised  from non vanishing replicas with shorter periodicity 
$(L_x, L_y, L_z)$. It is also simple to show that in the limit 
$n_x,n_y,n_z \to \infty$, the Hartree potential coincides with the 
exact integral expression in the physical volume $V$, where we solve the 
eigenvalue equations in Eq.(\ref{scfed}). 
Finally,  in order to have an efficient extrapolation for $n_x,n_y,n_z\to \infty$ at 
fixed volume, we have adopted a regularization  for the Coulomb potential 
when $r_I = r_J$ for open systems.
This is obtained by using an appropriate constant $C$,
 that is determined by imposing that the  
Hartree potential at $r=0$ coincides with the exact one 
for a Gaussian density  $n(\vec r)= \exp ( - r^2/2)$, namely
 $v_H(0)=\int dr^3  1/r \exp(-r^2/2)  = C+v \sum\limits_{ r_I \in V^A, r_J \ne 0 } \exp(-r_I^2/2)/r_I = 4 \pi $ in this case \cite{pbc}.
Obviously $C\to 0$ for $n_x,n_y,n_z\to \infty$ and therefore this regularization is only meant to accelerate the convergence but does not change the limit values of all the quantities computed with this SCF method.

\begin{figure}
\begin{center}$
\begin{array}{cc}
\includegraphics[width=0.5\columnwidth, angle=-00]{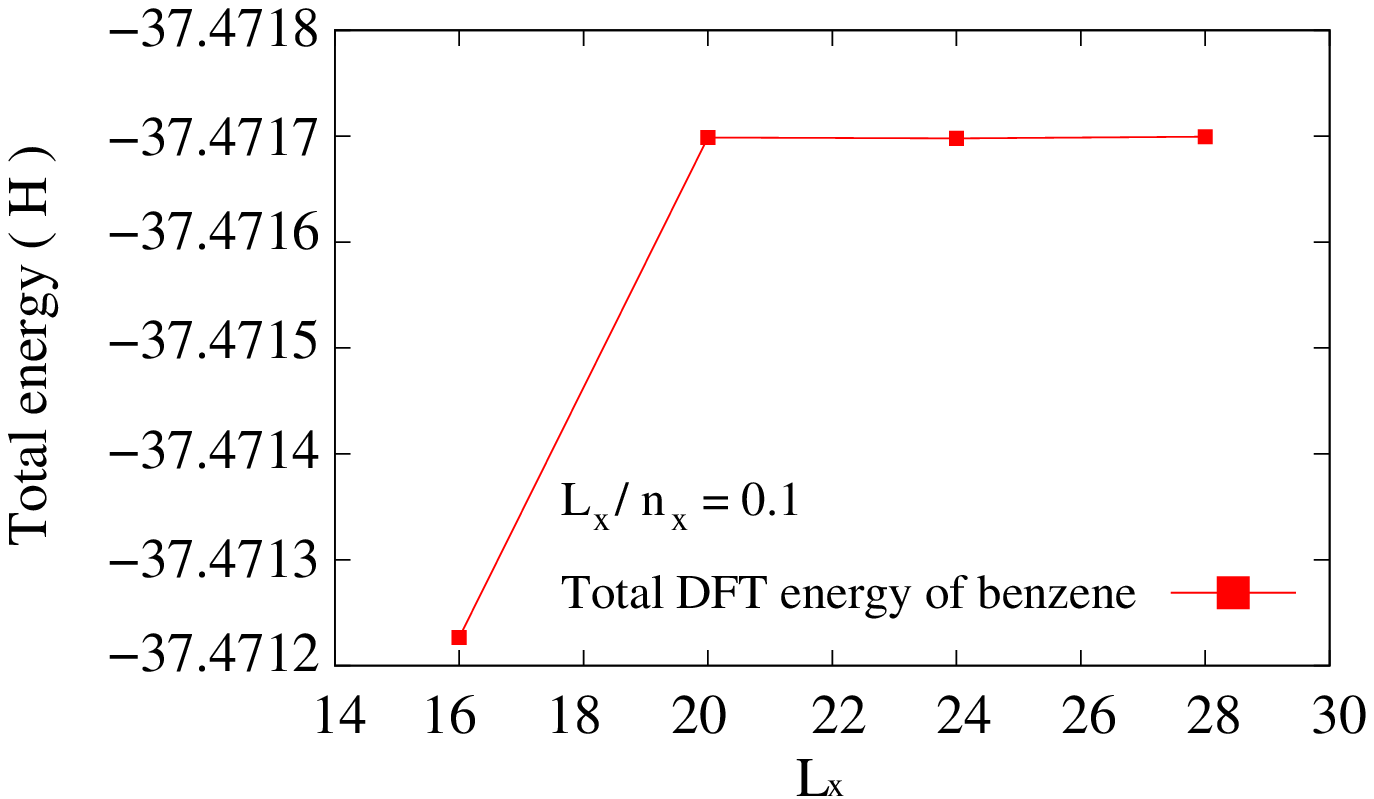} &
\includegraphics[width=0.5\columnwidth, angle=-00]{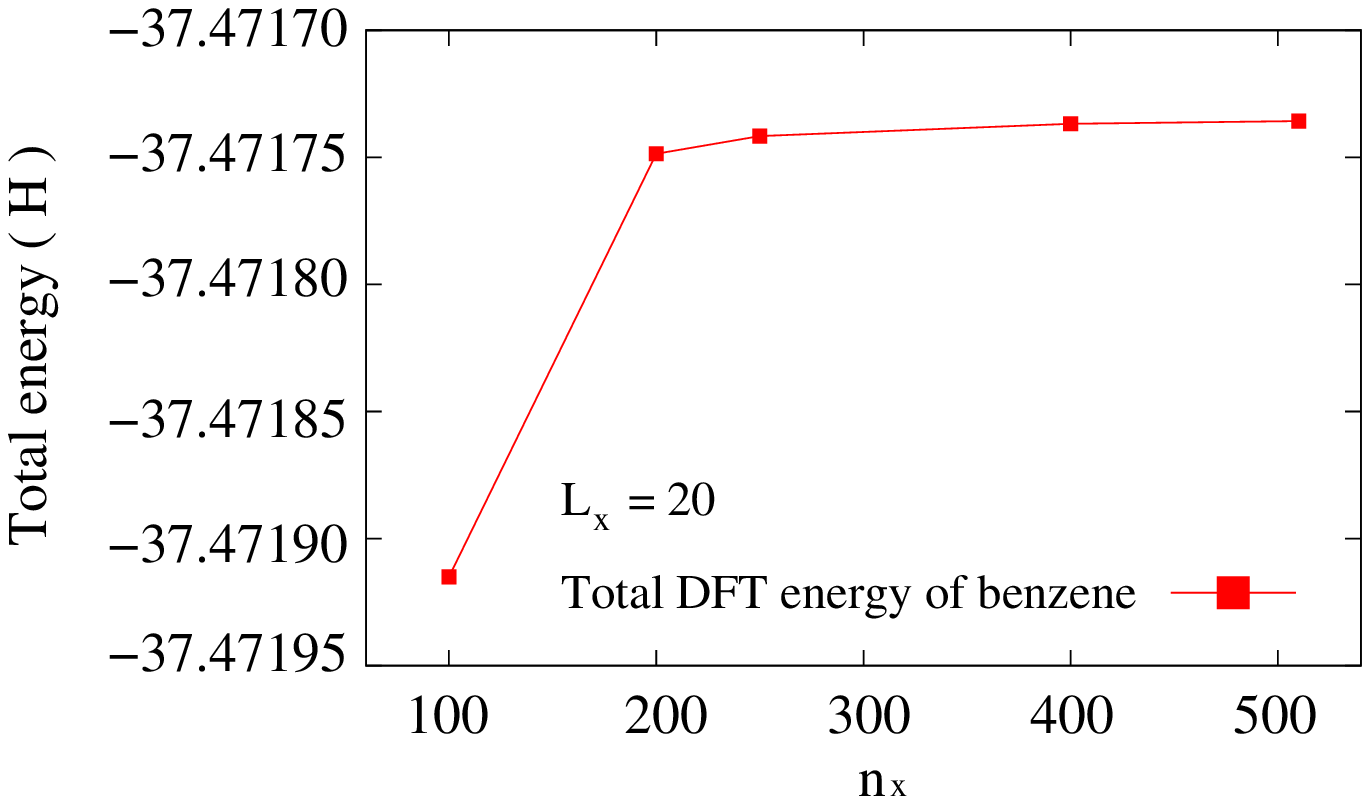}
\end{array}$
\end{center}
\caption{\small{Convergence of the total DFT energy of the benzene molecule as function of the box size $L_x$ (left), in $(Bohr)$,
 and the mesh size $n_x$ (right) using the $8s6p$ even-tempered basis set described in the text. }}
\label{box_converge}
\end{figure}

All the above simple technical improvements 
allow us  to have a very rapidly convergent 
calculation in $n_x, L_x  \to \infty$ that is otherwise almost impossible, 
especially for open systems.
In our application to the benzene molecule we have chosen a cubic box
 ($n_x=n_y=n_z$ and $L_x=L_y=L_z$) and the convergence vs the box  $L_x\to 
 \infty$ and mesh $n_x \to \infty$ size are displaied in Fig.~\ref{box_converge}. 

In Tab.~\ref{DFT_stability}, we show the same type of convergence  also 
for the exchange and correlation energies. 
It is clear that the rapid convergence of our method applies also to correlation functions and also that high quality 
results can be obtained with a reasonable computational effort. 

The way we  compute matrix elements on a grid 
is obviously much less efficient 
then  the standard way to use tabulated Gaussian 
integrals \cite{G09}. 
Nevertheless  the calculation we propose is quite simple and straightforward, obviously at some expense of computer time, but with the remarkable advantage 
that this method can be applied also to non-Gaussian orbitals, to periodic 
system, and that the 
accuracy of the calculation can be systematically controlled by increasing 
$n_x$ and the volume $V$.
This is particularly important when the condition number $s_{cond}$ becomes 
extremely large, and, for an accurate calculation, both overlap  and 
Hamiltonian matrix elements have to be calculated with much higher accuracy.
Moreover, for application to QMC computation, 
the time spent for a SCF calculation
represents only a negligible amount, so that it is not really important that 
this part of the calculation is fully optimized. 
The most important problem we have solved in this work is the stabilization
of the diagonalization routine for arbitrary large basis set and 
given (namely allowed by double precision arithmetic) accuracy  
in the calculation. This technique will be described in details in the 
next subsection.

\subsection{ Stable diagonalization routine} 
\label{dft_algorithm}

In this part  we describe our way to improve the 
accuracy of the SCF algorithm by means of a more stable numerical 
solution of the generalized eigenvalue equations given in Eq.\ref{scfed}. 
Basically the task is to compute in a stable way the eigenvectors of 
the Hamiltonian defined in a non-orthogonal basis of large linear dimension $N$.
The problem is that the overlap matrix $S$ in Eq.(\ref{sov}) may have a very large condition number 
and a straightforward diagonalization leads to inaccurate and often dirty eigenvectors.
In parallel computation there is also the further complication that 
it is difficult to preserve orthogonality (and accuracy) of eigenvectors
without having a huge buffer memory at disposal, as no efficient memory 
distribution  is possible for the  orthogonalization procedure.
Here we describe the algorithm of our diagonalization routine, 
that solves in a very simple and efficient way all the above 
computational issues:

\begin{itemize}
\item We apply first a diagonalization algorithm to the overlap matrix, 
based on the Householder tridiagonalization  and iterative Givens transform 
to the resulting tridiagonal matrix.
All these transformations are unitary and should preserve orthogonality 
of eigenvectors $ v_j^i$ for $i,j=1,\cdots, N$ in infinite precision arithmetic.
However this is not the case in practice since the matrix can be 
very ill-conditioned.
Therefore, in order to improve the stability of the calculation we 
follow standard procedures in numerical linear algebra\cite{book}. We 
disregard eigenvectors    with small eigenvalues  $s_i$ compared to the 
maximum one $s_N\le  N$, 
namely satisfying   $ s_i/s_N  <  \epsilon_{mach} $, where 
 $\epsilon_{mach}$ is an input parameter, whose minimum value  is around the 
relative machine precision ($\simeq 10^{-16}$ 
in double precision arithmetic)\cite{macheps}.
Neglecting small eigenvectors corresponding to small eigenvalues 
of the overlap matrix 
is justified from the fact that an eigenvector of the 
overlap matrix with nearly zero eigenvalue corresponds  to
a linear combination of normalized orbitals $\sum_j v^i_j \phi_j( \vec r) $, 
satisfying $ \sum_j  | v_j^i|^2 =1$,  
that has almost vanishing norm equal to $ \sqrt{s_i}$, 
i.e. the non-orthogonal basis of normalized orbitals is redundant and
this direction can be safely eliminated within an error $\sqrt{s_i}$.
Thus after neglecting all these singular directions we obtain a simple 
bound for the numerical error $\epsilon_{acc}$ expected when neglecting 
all these redundant directions:  
\begin{equation} \label{epsacc}
\epsilon_{acc} < Min_{i}  \left(  \sqrt{s_{i}}~|~  s_i/s_{N} \ge \epsilon_{mach}  \right)  \le \sqrt{ \epsilon_{mach} s_{N}} \le \sqrt{\epsilon_{mach} N}, 
\end{equation} 
where in the last inequality we have used that $s_N \le N$, as shown  
before.

In this step we do not require that the 
diagonalization has produced really  orthogonal eigenvectors 
but only that the eigenvalues have the right order of magnitude,
the normalization of eigenvectors is correct,  
and that the diagonalization routine 
has produced linearly independent eigenvectors, 
properties that are easily 
satisfied even for extremely singular overlap matrices.
Then we define non-singular directions:
\begin{equation} \label{newbasis}
e^i_j = { 1 \over \sqrt{s_i}}  v^i_j ~ {\rm for }~  s_i/s_{N}  \ge \epsilon_{mach} 
\end{equation}

In this basis, 
the overlap matrix $ \tilde s_{i,j} =\langle e_i | S | e_j \rangle $ should be equal to the 
identity in infinite precision arithmetic, namely a matrix with minimum  
condition number $s_{cond}=1$. 
Thus it turns out that, in finite precision 
arithmetic, by recomputing $\tilde s_{i,j} = \sum_{k,l} e^i_k S_{k,l} e^j_l $
we obtain a well-conditioned matrix that can be diagonalized again 
with high accuracy:
\begin{equation}
 \tilde s_{i,j} =\sum_k  \tilde s_k \tilde v^k_i \tilde v^k_j
\end{equation}
and therefore now, analogously to the previous case, 
we can define   directions
 $$\tilde e^i_k = { 1 \over \sqrt{ \tilde s_i}}  \tilde v^i_k $$
that remain safely saorthonormal even in finite precision arithmetic.
Finally  we store the global transformation from the original basis to the 
new one
\begin{equation}
 U_{i,j} = \sum_k e_j^k  \tilde e^i_k  
\end{equation}
In this basis the generalized eigenvalue 
equation (\ref{scfed}) turns in a standard diagonalization 
of a very well-conditioned Hamiltonian matrix $\tilde H =U H U^{\dag}$ 
because its spectrum corresponds to the physical spectrum of the original 
Hamiltonian. 
\item At each iteration we recompute the Hamiltonian matrix $\tilde H$ 
in this new basis. 
\item Then apply again our diagonalization routine. 
\item Go back to  the original basis $W^i_j = \sum_k U_{j,k} \tilde  W^i_k $ for the eigenvectors  $W^i_j$ ($\tilde W^i_k$) of $H$ ($\tilde H$),  computed  at each iteration to implement self consistency and write the final molecular orbitals 
in the localized atomic basis set.
\end{itemize}
 
 The above  scheme  allows us to obtain a total energy accuracy that, at least in the  examples studied in this work, is below $0.1mH$, namely one order of magnitude smaller than the typical target chemical accuracy.
This is achieved in an automatic way, by using  
$\epsilon_{mach} \simeq 10^{-16}$.
Namely, even when the basis used is extremely redundant  for $N \to \infty$, 
the simple algorithm, that  we have previously described,  allows us  to obtain accurate total energies. 
This solution is general in the following sense: with the present algorithm 
it is in principle possible to work  
with arbitrarily large value of the basis dimension $N$, 
namely even with an over-complete basis set, and, after removing  redundant directions as described in the first step of the algorithm, we can  work with a well-conditioned orthonormal basis, obtained after the second diagonalization, 
and that provides for large $N$ converged results within the 
numerical accuracy $\epsilon_{acc}$ 
possible with the available numerical precision.  
Since, as shown before in Eq.(\ref{epsacc}), $\epsilon_{acc}$ increases very slowly with $N$,
namely at most as  $\propto \sqrt{N}$, we do not expect accuracy problems 
with  basis sets much larger than the ones  used in the examples 
presented here, even within double precision arithmetic.

\begin{figure}
\begin{center}
\includegraphics[width=0.5\columnwidth, angle=-00]{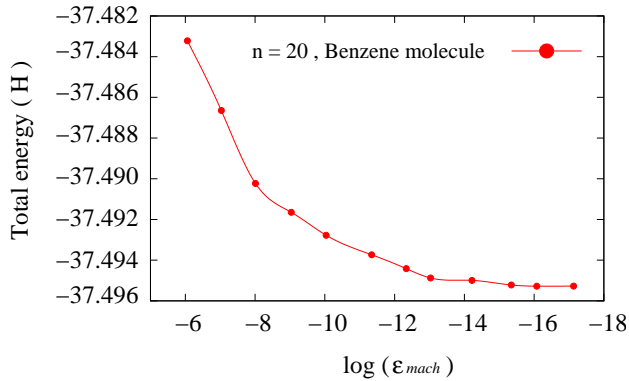}
\end{center}
\caption{\small{Total DFT energy of benzene molecule in $n=20$ case as a function
 of our regularization cutoff $\epsilon_{mach}$. }}
\label{epsover}
\end{figure}

As shown in Fig.~\ref{epsover}, the accuracy can be also controlled by decreasing   
 the value of $\epsilon_{mach}$, a  lower  value leads to a   more accurate 
 calculation, until one reaches a threshold below  the relative machine 
precision  when 
instabilities occurs in the diagonalization because the numerical accuracy 
of finite precision arithmetic is simply not enough. 
Better accuracies  could be  in principle possible only with more accurate operations,  e.g. by using quadruple precision. 
Nevertheless, as anticipated, this algorithm provides a reasonable accuracy in the total energy with standard, 
and usually much more efficient, double precision arithmetic. 

\begin{table}
\caption{Total, exchange and correlation energies of DFT calculation of 
the benzene molecule using $24s22p10d6f$ basis set. A cubic box ($L_x = L_y = L_z $) is used. The given 
values of energies and box sizes are in (H) and $(Bohr)$, respectively. }
\label{DFT_stability}
\begin{center}
\begin{tabular}{|c|c|c|c|c|}
\hline\hline
\hline
$n_x$  & $L_x$ & Total energy  & Exchange energy  & Correlation energy   \\
\hline
160 & 16 &  -37.5404624 & -10.6865190 & -1.5922693  \\
200 & 20  & -37.5400762 & -10.6867874 & -1.5922738 \\
240 & 24  & -37.5400821 & -10.6867701 & -1.5922721 \\
400 & 20 & -37.5400657 & -10.6867885 & -1.5922732 \\
\hline\hline
\end{tabular}
\end{center}
\end{table}

\subsection{Correlated variational wave function for QMC calculations}

In this part we describe our WF which has been used in QMC calculations.
The usual trial wave function (WF) used in QMC calculation is the product of an
antisymmetric part and a Jastrow factor. The antisymmetric part is a single Slater determinant, while the Jastrow factor is a bosonic 
many body function which accounts for the dynamic correlations 
in the system. Our Slater determinant is obtained with
$N_{el}/2$ ($N_{el}$ is the total number of electrons in the system) doubly occupied molecular orbitals $\psi_j(r)$, expanded in atomic orbitals as described in Eq.(\ref{defpsi}).
The molecular orbitals are obtained from the
 self consistent DFT-LDA calculations explained in the previous section.
 The Jastrow factor takes into account the electronic correlation between two 
electrons and  is conventionally split into an homogeneous interaction $J_2$ depending on the relative 
distance between two electrons,  and a non homogeneous contribution depending on the positions of one or two atoms, 
$J_{3}$ and $J_{4}$ respectively. It also contains a one particle term $J_1$, that is important to compensate the 
change in the one particle density induced by $J_2$, $J_3$ and $J_4$, as well as to satisfy the electron-ion cusp conditions.
 The one- and two-body terms $J_1$ and $J_2$ are defined by the following equations: 

\begin{eqnarray}
\label{j1}
J_1=\exp{\big[\sum_{ia}-(2Z_a)^{3/4}u(Z_a^{1/4}r_{ia})+\sum_{ial}
g_l^a \chi_{al}^J(\vec{r}_i)\big]},
\end{eqnarray}
and 
\begin{eqnarray}
\label{j2}
J_2=\exp{[\sum_{i<j}^{}u(r_{ij})]},
\end{eqnarray}
where $i,j$ are indices running over the electrons, and $l$ runs over different single particle orbitals $\chi_{al}^J$ 
centered on the atomic center $a$. $r_{ia}$ and $r_{ij}$ denote electron-ion and 
electron-electron distances respectively. The corresponding cusp conditions are fixed by 
the function $u(r)=F[1-\exp(-r/F)]/2$ (see e.g. Ref.~\cite{rocca}).
$g_l^a$ and $F$ are optimizable variational parameters. 
The three and four-body Jastrow $J_{3} J_{4}$ are  given by:
\begin{equation}
\label{jastrow}
J_{3} J_{4}(\vec R)=
\exp\left(\sum \limits_{i<j} f(\vec r_i ,\vec r_j)\right), 
\end{equation}
with $f (\vec r_i ,\vec r_j)$, being  a two-electron coordinate 
function that  can be expanded into the same single-particle basis 
used for $J_{1}$: 
\begin{eqnarray}\label{3bjsp}
f(\vec{r}_i,\vec{r}_j)=
\sum_{ablm}^{} g_{lm}^{ab}\,\chi_{al}^{J}(\vec{r}_i)
\chi_{bm}^{J}(\vec{r}_j),
\end{eqnarray}
with $g_{lm}^{ab}$ optimizable parameters. Three-body (electron ion electron) correlations are described by the diagonal matrix elements 
$g^{aa}$, whereas four-body correlations (electron ion electron ion) are described by matrix elements with $a\ne b$.

The  exhaustive and complete expression of the  Jastrow factor  $J(\vec R) = J_1(\vec R) J_2 (\vec R) J_3(\vec R) J_4(\vec R) $ 
that we adopt in this work allows us to take into account not only weak electron-electron interactions, 
but it is also extremely effective for suppressing  higher energy configurations occurring when electrons are too close.

\section{Results and discussion}

In this section we show the remarkable convergence and stability properties 
 of our method for the calculation of the total and atomization energies of the benzene 
molecule in the CBS limit.
To this purpose we consider an atomic basis with $l_{max} \le 1$, 
namely with only $s$ and $p$ type of orbitals allowed, and show that it is 
possible to converge to the $n\to \infty$ case even   
when, for large $n$, too many orbitals of the same angular momentum 
become highly redundant and  are difficult to treat with standard methods.
 
\begin{figure}
\begin{center}$
\begin{array}{cc}
\includegraphics[width=0.5\columnwidth, angle=-00]{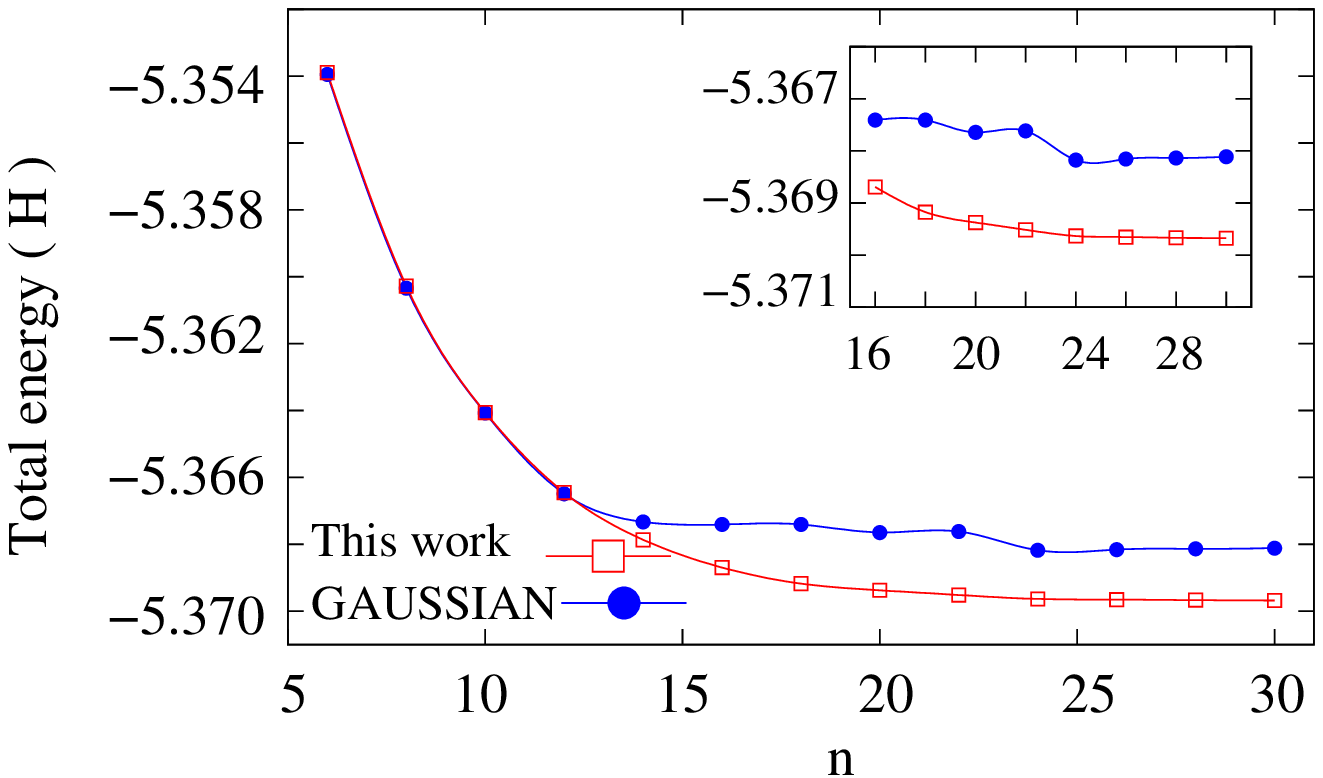} &
\includegraphics[width=0.5\columnwidth, angle=-00]{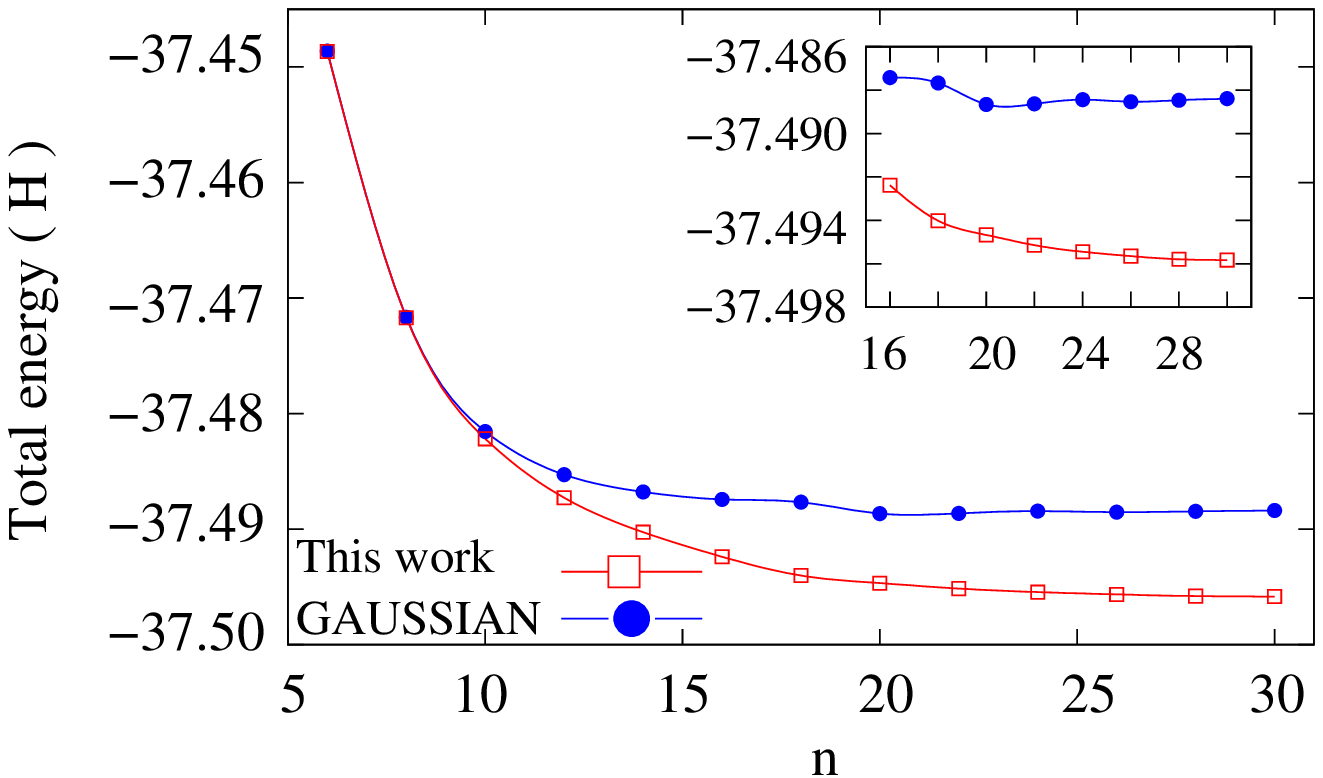} \\ 
\includegraphics[width=0.5\columnwidth, angle=-00]{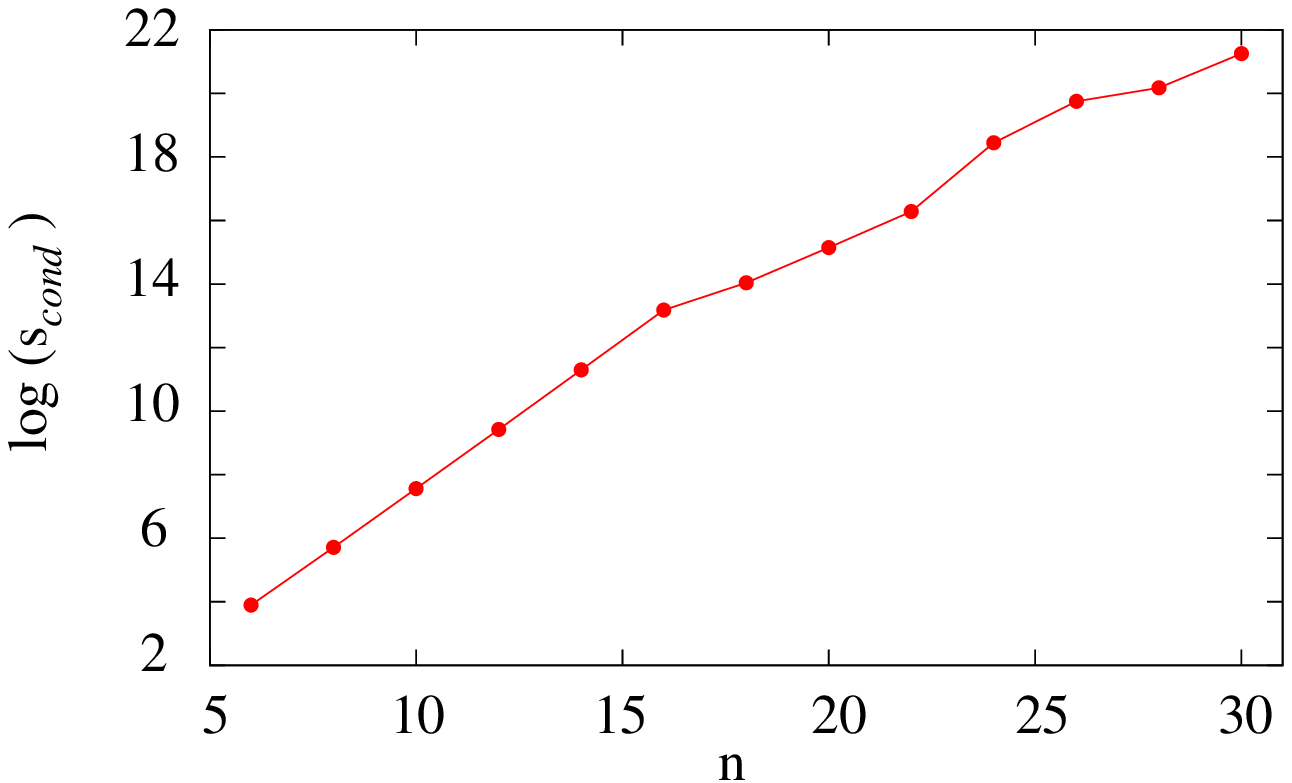} &
\includegraphics[width=0.5\columnwidth, angle=-00]{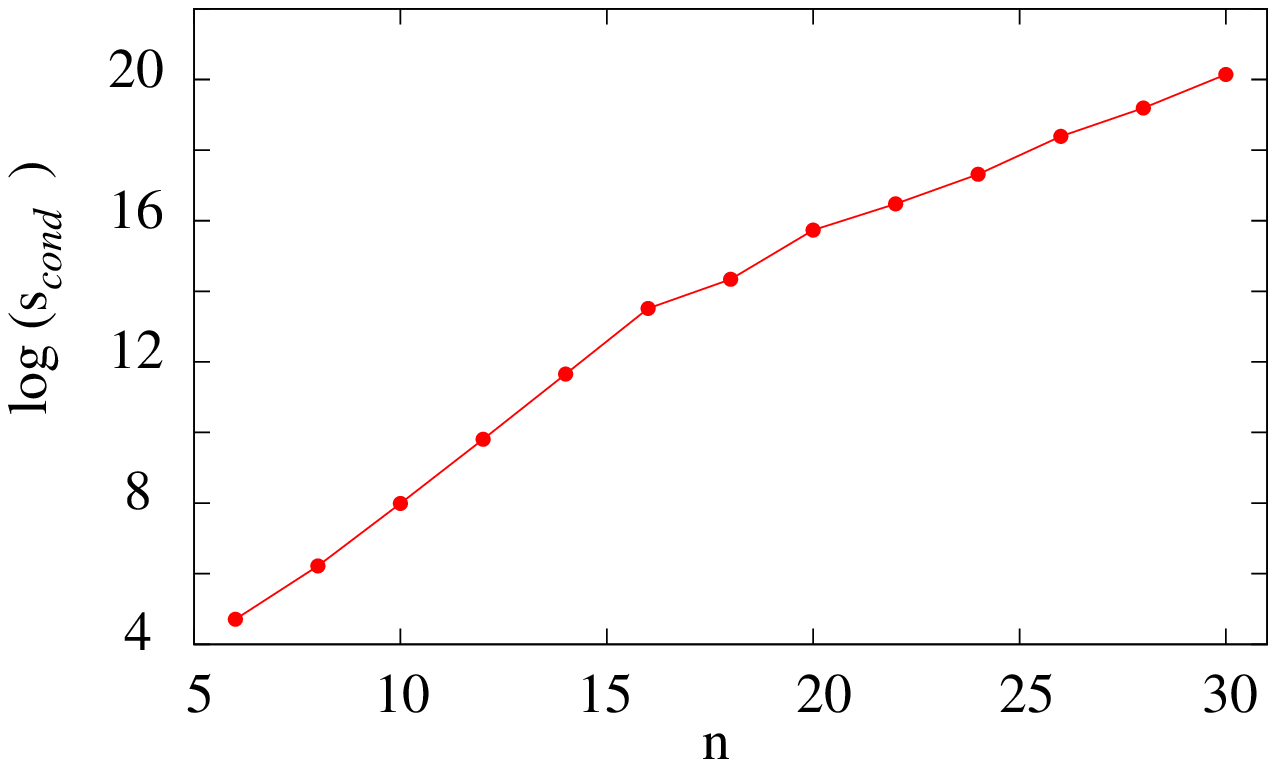}
\end{array}$
\end{center}
\caption{\small{Upper panels: convergence of the total DFT energy of the carbon atom (left) and the benzene molecule (right) using  an atomic 
 basis set containing only $s$ and $p$ angular momenta. 
 Results obtained by GAUSSIAN09 are also shown for comparison. Lower panel,  Logarithm of the condition number $s_{cond}$ 
 of the overlap matrix for the carbon atom (left) and the benzene molecule (right). }}
\label{DFT_C_benzene}
\end{figure}

In Fig.~\ref{DFT_C_benzene} we compared our total DFT energies of the carbon 
 atom and the  benzene molecule  with the ones obtained with the 
GAUSSIAN09 package \cite{G09}.
We have used both for  GAUSSIAN09 and our DFT algorithm exactly the same basis sets and pseudopotentials, treated with maximum accuracy in the angular integration, and 
therefore with negligible error, as well as the same Slater exchange and correlation 
functional with  the standard Perdew-Zunger parameterization \cite{perdew81}.

\begin{table}
\caption{\small{Total DFT energy for the benzene molecule as a function of the
highest angular momentum of the atomic basis.
Calculations were done with pseudopotentials \cite{filippipseudo}
both for the hydrogen and carbon atoms  
at the experimental equilibrium positions, while  
 for the hydrogen we have used a  $3s2p$ basis.}}  
\label{benzene_dft_energies}
\begin{center}
\begin{tabular}{|c|c|c|c|c|}
\hline\hline
\hline
$L$  &  Basis & C-composition & number of primitive Gaussian & Energy (H) \\
\hline
1 & sp limit & 24s22p & 594 & -37.4952567 \\
2 & spd limit  & 24s22p10d & 894 & -37.5357514 \\
3 & spdf limit  & 24s22p10d6f & 1146 & -37.5400821 \\
4 & spdfg limit & 24s22p10d6f2g & 1254 & -37.5416815 \\
\hline\hline
\end{tabular}
\end{center}
\end{table}

For the small basis sets ($n_s=6,8$) we are in excellent 
agreement with GAUSSIAN09.
However, with large basis sets ($n_s>8$) there is a clear 
difference between our results and the GAUSSIAN09 ones. 
We do not know exactly 
 what is the reason of this discrepancy. We just report that,  as it is shown 
in the lower panels of Fig.~\ref{DFT_C_benzene}, the condition number 
$s_{cond}$, is increasing quite rapidly with the dimension $N$ of the basis 
set, and the discrepancy between our results and GAUSSIAN09 is evident 
when the condition number becomes larger than $\simeq 10^{8}$.
Moreover since DFT is a variational method, it should be clear that 
for a given basis set the method that provides the lowest value of the 
functional should be the most accurate, provided the value of the functional 
can be calculated accurately.
Indeed we have verified that 
a large  condition number $s_{cond} \sim 10^{17} $ affects only  
the self consistent step (in our case the diagonalization in  
Eq.\ref{scfed}) but allows to compute  
the value of the functional without particular problems \cite{checkdft}.
Therefore we can safely state that our Kohn-Sham molecular orbitals are very well converged, whereas standard methods suffers to work already 
with condition number larger than $10^8$.

\begin{figure}
\begin{center}
\includegraphics[width=0.5\columnwidth, angle=-00]{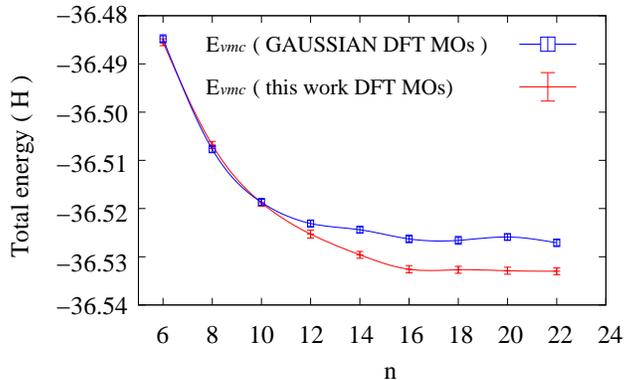}
\end{center}
\caption{\small{VMC computation of the total energy  $E_{vmc}$
of the Slater  determinant obtained with Kohn-Sham molecular orbitals. 
For comparison we show also the results obtained with GAUSSIAN09.}}
\label{E_vmc_noJ}
\end{figure}

As a matter of fact our DFT energy for the benzene molecule converges 
to a very good value $-37.4951453 (H)$, (e.g. the GAUSSIAN09 
result in the same basis is  $-37.4886324 (H)$).
The quality of our very well converged 
molecular orbitals is also evident when we 
compute the expectation value 
of the energy  
of the corresponding Slater determinant  
by using the standard 
variational Monte Carlo technique to compute the energy expectation values 
(see Fig.~\ref{E_vmc_noJ}).
Though it is clearly inefficient to use VMC to compute energy expectation 
values of uncorrelated wave function, it is useful to use this method
in this case because it does not require 
the explicit evaluation of the matrix $S$ with very large condition number. 

Moreover the study of the $n=20$ case as a function of our regularization 
cutoff $\epsilon_{mach}$, displaied in Fig.~\ref{epsover},
 clearly shows that we can reach a sufficient 
accuracy ($\sim 0.1mH$) in the total energy 
even in the large basis set limit, and that therefore 
the results for $n=24$ should be considered well converged, namely 
close to the $n\to \infty$, $l_{max}=1$ case. 

\begin{figure}
\begin{center}
\includegraphics[width=0.5\columnwidth, angle=-00]{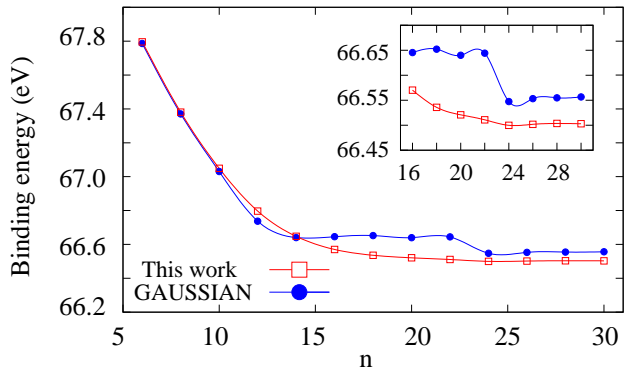}
\end{center}
\caption{\small{DFT atomization energy of the benzene molecule calculated  
by using a basis set containing only $s$ and 
$p$ atomic orbitals ($l_{max}=1$ limit). Results obtained by GAUSSIAN09 are 
also shown for comparison.}}
\label{binding_Tur_Gau}
\end{figure}

It is also particularly interesting to show how the atomization energy 
converges in this case. In Fig.~\ref{binding_Tur_Gau} our results 
clearly indicate that the so called basis superposition error is 
very important in this case and monotonically disappears only for large 
$n$. 

Finally we study the convergence of the total energy as 
a function of the maximum angular momentum $l_{max}$ of the 
atomic basis. 
DFT energies are shown in Tab.~\ref{benzene_dft_energies}.
Though the difference between $l_{max}=4$ (spdfg) and $l_{max}=3$ (spdf) 
is larger than $1 mH$, this table shows that the convergence with 
$l_{max}$ is quite rapid, as
each time $l_{max}$ is increased by  a unit  the accuracy improves by more than 
 a factor $3$, and therefore for $l_{max}=4$ 
we should be very close to the CBS limit, 
well within $1 mH$ accuracy in the total energy.

In the following  we show the importance of being close to the CBS limit in QMC 
calculations obtained either by optimizing the Jastrow factor over the 
Slater determinant defined by the DFT Kohn-Sham molecular orbitals 
or by full optimizing both the Jastrow and the molecular orbitals 
with  the method described in  Ref. \cite{Mari}, starting from 
the  former initial wave function. 
We indicate  in the following the first wave function by $J-DFT-WF$, whereas 
the latter one will be denoted by $J-OPT-WF$. 

\begin{figure}
\begin{center}
\includegraphics[width=0.5\columnwidth, angle=-00]{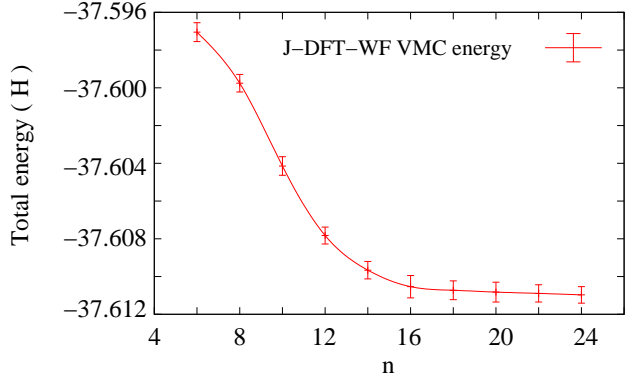}
\end{center}
\caption{\small{VMC total energy using the $J-DFT-WF$, defined in the text,  for the benzene molecule as a function of the 
number of $s$ and $p$ orbitals.
 We have used DFT molecular orbital as a determinant part of trial wave function and only the Jastrow factor was optimized. }}
\label{VMC_JDFT_converge_sp}
\end{figure}

A $4s3p2d$ ($1s1p$) uncontracted Gaussian basis set was used 
for expanding the Jastrow factor around each carbon (hydrogen) atom. 
Fig.~\ref{VMC_JDFT_converge_sp} shows the convergence of VMC total energy of the benzene molecule with $J-DFT-WF$, which
an atomic basis used for the Slater determinant with $l_{max}=1$. 
This picture shows that the presence of the Jastrow factor  
improves the  convergence to the $n\to \infty$ limit.
For instance the DFT total energy converges when the basis contains more 
than   $24s$ and $22p$. Instead, by using the Jastrow factor, 
clear convergence is reached  already for a  $16s14p$ basis set.
For fixed  maximum angular momentum of the atomic basis  $l_{max}$ in the 
Slater determinant, we have carried out VMC calculations of the total energy of the benzene molecule.
As shown in Tab.~\ref{totalVMC_JDFT_WF} 
the total energy difference between the $l_{max}=1$ and the $l_{max}=2$ cases 
 is about $0.75 (eV)$, while this difference for  $l_{max}=2$ and $l_{max}=3$
shrinks to  $0.25 (eV)$. Moreover our VMC results suggest that it is not 
necessary to  add $g$ orbital to the basis set, for an accuracy  
smaller than  $0.005 (eV)$. 

\begin{table}
\caption{\small{VMC total energy (H) of benzene molecule by using  
the $J-DFT-WF$ defined in the text. }}
\label{totalVMC_JDFT_WF}
\begin{center}
\begin{tabular}{|c|c|c|c|}
\hline\hline
\hline
$l$  &  Basis & C-composition &  Energy (H) \\
\hline
1 & sp limit & 24s22p & -37.6111(4) \\
2 & spd limit  & 24s22p10d & -37.6386(4) \\
3 & spdf limit  & 24s22p10d6f & -37.6478(4) \\
4 & spdfg limit & 24s22p10d6f2g & -37.6480(4) \\
\hline\hline
\end{tabular}
\end{center}
\end{table}

The VMC atomization energy of the benzene molecule within  the $J-DFT-WF$ for fixed  $l_{max}$ 
is reported in Tab.~\ref{bindingVMC_JDFT_WF}. The estimated exact atomization  energy of the benzene molecule is $59.33(3) (eV)$ \cite{parthiban}, by neglecting inner-shell correlation, that we assume to be negligible with  the pseudopotentials we have used \cite{umrigarprl}.
Therefore,  by optimizing the Jastrow factor, one can get almost exact atomization  energy
as long as a large basis set is considered for the DFT Slater determinant.

\begin{table}
\caption{\small{VMC atomization  energy of the 
benzene molecule obtained with the  $J-DFT-WF$ defined in the text.}}
\label{bindingVMC_JDFT_WF}
\begin{center}
\begin{tabular}{|c|c|c|c|}
\hline\hline
\hline
$l$  &  Basis & C-composition &  Binding Energy (eV) \\
\hline
1 & sp limit & 24s22p & 58.38(4) \\
2 & spd limit  & 24s22p10d & 59.12(4) \\
3 & spdf limit  & 24s22p10d6f & 59.37(4)  \\
4 & spdfg limit & 24s22p10d6f2g & 59.37(4) \\
\hline\hline
\end{tabular}
\end{center}
\end{table}

One of the main outcome of our work, is that the DFT Slater-determinant is a very good  input for QMC calculations,
provided the basis used is sufficiently large. 
In fact by full optimization of both the Jastrow and the determinantal
  parts of the WF and  by using a large basis for the Slater determinant $24s22p10d6f$, 
the VMC total energy of the benzene molecule is $-37.6491(3) (H)$.
 Hence, the difference between the VMC total energy
using the $J-DFT-WF$ and the $J-OPT-WF$ is very small  $0.035 (eV)$ and  this  shows that, 
by optimizing the determinantal part of the WF, the total energy
improves only by a small amount and does not appreciably change the atomization  energy of benzene, from $59.37(4) (eV)$ to $59.41(3) (eV)$. 
Remarkably from a DFT-LDA  atomization  energy that is completely wrong by $\sim 10 (eV)$,   
 we can obtain an almost exact atomization energy 
using the $J-DFT-WF$ wave function with  
the Kohn-Sham molecular orbitals  
obtained with  the energetically poor  DFT-LDA method.

A much different behaviour is obtained when the wave function is 
 fully optimized   within a small basis set. Indeed, 
we have applied full optimization on the smallest basis set $6s4p$, and the 
VMC total energy using the $J-OPT-WF$ is $-37.6384(5) (H)$, which is
 $ 1.12 (eV)$  below the corresponding $J-DFT-WF$ energy. This energy 
gain is more than one order of magnitude larger than the one obtained  
in the previous case. Therefore we conclude that the Kohn-Sham molecular  
orbitals are very accurate only when a sufficiently large basis is used in the 
DFT calculation.

In this limit, in order to show the quality of our variational wave functions,
we have carried out LRDMC \cite{lrdmc} calculations using the $J-DFT-WF$ and the $J-OPT-WF$. 
The LRDMC total and atomization  
energies of the benzene molecule are shown in Tab.~\ref{LRDMC}. Though the 
LRDMC improves the VMC total energy of the benzene molecule by 
about  $\simeq 1.7 (eV) $, the atomization energy remains  unchanged within the statistical errors.

\begin{table}
\caption{\small{LRDMC total and atomization energies of the benzene molecule
 obtained by the  $J-DFT-WF$ and the $J-OPT-WF$, described in the text.}}
\label{LRDMC}
\begin{center}
\begin{tabular}{|c|c|c|}
\hline\hline
\hline
 &  $J-DFT-WF$ & $J-OPT-WF$  \\
\hline
Total energy (H) & -37.7111(5) & -37.7128(4) \\
Binding energy (eV) & 59.41(5)  & 59.45(4) \\
\hline\hline
\end{tabular}
\end{center}
\end{table}

\section{Conclusion}
We have introduced a very simple method to make accurate and 
 systematically converging SCF calculations with localized basis sets 
of increasing dimension. 
This work shows that the use of a large basis set may be 
extremely important for accurate calculations.
It is possible that  our method could be relevant also for defining 
more efficient electronic structure  packages that are free of any limitation 
about basis set dimension.
Preliminary application of the method to periodic systems are also 
extremely encouraging\cite{silicon}, because one can work also in this 
case with a localized basis set, with convergence properties similar to 
plane wave DFT methods.
For large extended systems the condition number of a GTO localized basis set increases quite rapidly with the system size  but 
we have tested in Silicon with a supercell containing up to 256 atoms (i.e. 1024 valence electrons) that  our algorithm remains stable  
for fixed  choice of the parameter $\epsilon_{mach} \simeq 10^{-16}$ used to remove the singular directions. 
Although it impossible to obtain  the error of the finite basis used 
(8s6p4d per atom)  because the CBS limit is computationally too expensive for large number of electrons, by comparing our results with standard plane wave methods it turns out that  our  accuracy should remain constant for the energy per atom. This is expected from general grounds, since for extended systems the condition number of a localized 
basis set,  defined by a fixed number of orbitals per atom,  should saturate in the infinite volume limit, when orbitals corresponding to atoms that are very far apart remain orthogonal, because they do not overlap.
As we have already remarked before better accuracy -probably  necessary for 
large extended systems when for instance the chemical accuracy in the total energy is required- can 
be in principle possible with much smaller $\epsilon_{mach}$, that can be 
used  only with a more accurate arithmetic (e.g. quadruple precision).

Moreover we have shown that,  for QMC applications,  our method is  extremely useful,  because 
only in the large basis set limit the output molecular orbitals of our new  SCF
calculation define an extremely accurate Slater determinant, that essentially, 
does not need to be optimized.
This work also highlights a remarkable property of the DFT method, namely that  
, rather surprisingly, the Kohn-Sham molecular orbitals are rather robust and stable  in the large basis set limit, and do not seem to be very much sensitive to  the accuracy of the functional used. In the benzene example for instance, the 
accuracy in the atomization energy with the LDA functional used was very poor 
with an error of about $\simeq 10eV$, whereas when the same Kohn-Sham orbitals 
are used for standard QMC calculation, they provide almost optimal  results,
not only compatible with experiments, but also stable against further 
optimization of the energy in presence of the Jastrow factor.

\acknowledgements
This work is partially supported by MIUR-COFIN07, and CINECA. 
One of us (SA) acknowledges useful discussions with M. Parrinello.


%
  

\begin{thebibliography}{99}
\bibitem{pbc} On the other hand for periodic systems, standard Ewald summations 
are employed and the  
 Coulomb potential $v_c(q)= 4 \pi/ q^2 = v_L(q)+v_S(q)$ is decomposed in a long range $v_L (q) = 4 \pi \exp ( -q^2/4 k^2)/q^2$  
 and a short range part $v_L(q) = v_c(q)-v_L(q)$, where $k$ is the Ewald 
constant.
The short range part has an exact expression in real space:
\begin{equation}
v_S(r) = {\rm erfc}(k r)/r
\end{equation}
and is used for computing efficiently the electron-ion and ion-ion 
Coulomb energy in the real space mesh.
Instead in the electron-electron Coulomb energy it is convenient to employ  
the Fourier convolution with  a slightly different 
Coulomb potential  $\tilde v_c(q)=v_L(q)+\tilde v_S(q)$  
where $\tilde v_S(q)$ is the discrete Fourier transform  for the short range part $\tilde v_S(q)= v 
\sum\limits_{r_I \ne 0, r_I \in V} e^{ - i q r}  v_S (r)+ C$. 
Indeed, 
it is convenient to use $\tilde v_c(q)$ in place of $v_c(q)$ in the Fourier 
convolution in order to be consistent with the  calculation of 
the electron-ion and ion-ion contribution to the Coulomb energy.
Moreover, analogously to the open system case, in order to obtain a much faster 
convergence the constant $C$ is determined by requiring that:
$$\tilde v_S(q=0)= \pi/k^2 $$, namely the  exact  
value of  the integral for $n_x,n_y,n_z \to \infty$.
\bibitem{macheps} The relative machine precision $\epsilon$ 
 is the minimum number such that $1+\epsilon \ne 1 $ in  
finite precision arithmetic.
\bibitem{checkdft}
In fact we have taken the GAUSSIAN09 molecular orbitals for the  $n=20$ case, and used them to evaluate the 
DFT functional value with our algorithm, finding perfect agreement with 
the GAUSSIAN09 DFT-energy.
\end{thebibliography}
\end{document}